\documentclass[10pt, a4paper, twocolumn, twoside]{IEEEtran}

\usepackage{pdfsync}

\usepackage{cite}
\usepackage{enumerate}
\usepackage{graphicx}
\usepackage[cmex10]{amsmath} 
\usepackage{amssymb}
\usepackage{mathtools}
\usepackage{xspace}
\usepackage{subfigure}
\usepackage[lined,linesnumbered,ruled,commentsnumbered]{algorithm2e} 
\usepackage{colonequals}
\usepackage[mathscr]{euscript}
\usepackage{fixltx2e}    
\usepackage{amsthm}
\usepackage{mathrsfs}
\usepackage{hyperref}
\usepackage[belowskip=-25pt,aboveskip=0pt]{caption}
 
\usepackage{epsfig}
\usepackage{epstopdf}

\usepackage{tikz}
\usepackage{pgfplots}

\usetikzlibrary{arrows,shapes,backgrounds,plotmarks,positioning}

\pgfplotsset{compat=newest}                         
\pgfplotsset{plot coordinates/math parser=false}
\newlength\figureheight
\newlength\figurewidth

\DeclareMathOperator*{\argmax}{arg\,max}
\DeclareMathOperator*{\argmin}{arg\,min}

\newtheorem{theorem}{Theorem}

\newtheorem{definition}{Definition}
\newtheorem{corollary}{Corollary}
\newtheorem{proposition}{Proposition}

\newtheorem{remark}{Remark}

\newcommand{\Set}[1]{\{#1\}}

\newcommand{\X}{\mathcal{X}}

\newcommand{\Y}{\mathcal{Y}}
\renewcommand{\S}{\mathcal{S}}
\newcommand{\eps}{\epsilon}
\newcommand{\Xepsc}{\X_\epsilon^c}

\begin{document}

\title{On Properties and Optimization of Information-theoretic Privacy Watchdog\vspace{-1mm}}

\author{Parastoo Sadeghi, Ni~Ding, and Thierry~Rakotoarivelo\thanks{P.~Sadeghi is with the School of Engineering and Information Technology, The University of New South Wales, Canberra. Her  work  is partly supported by the Data61 CRP: IT-PPUB. N.~Ding is with the School of Computing and Information Systems, University of Melbourne, Melbourne. T.~Rakotoarivelo is with Data61, Commonwealth Scientific and Industrial Research Organisation, Australia. Emails: \texttt{p.sadeghi@unsw.edu.au, ni.ding@unimelb.edu.au, thierry.rakotoarivelo@csiro.au.} The first two authors contributed equally to the paper.}}

\maketitle

\begin{abstract}
We study the problem of privacy preservation in data sharing, where $S$ is a sensitive variable to be protected and $X$ is a non-sensitive useful variable correlated with $S$. Variable $X$ is randomized into variable $Y$, which will be shared or released according to $p_{Y|X}(y|x)$. We measure privacy leakage by \emph{information privacy} (also known as \emph{log-lift} in the literature), which guarantees mutual information privacy and differential privacy (DP). Let $\Xepsc \subseteq \X$ contain elements n the alphabet of $X$ for which the absolute value of log-lift (abs-log-lift for short) is greater than a desired threshold $\eps$. When elements $x\in \Xepsc$ are randomized into $y\in \Y$, we derive the best upper bound on the abs-log-lift across the resultant pairs $(s,y)$.  We then prove that this bound is achievable via an \emph{$X$-invariant} randomization $p(y|x) = R(y)$ for $x,y\in\Xepsc$.  However, the utility measured by the mutual information $I(X;Y)$ is severely damaged in imposing  a strict upper bound $\eps$ on the abs-log-lift. To remedy this and inspired by the probabilistic ($\eps$, $\delta$)-DP, we propose a relaxed ($\eps$, $\delta$)-log-lift framework. To achieve this relaxation, we introduce a greedy algorithm which exempts some elements in $\Xepsc$ from randomization, as long as their abs-log-lift is bounded by $\eps$ with probability $1-\delta$. Numerical results demonstrate efficacy of this algorithm in achieving a better privacy-utility tradeoff.\vspace{-3mm}
\end{abstract}

\section{Introduction}
\subsection{Motivation and Background}
Most businesses collect increasingly large amounts of data about their customers or users. Such data has become a critical business asset in itself. For example, when analyzed using machine learning algorithms, it enables targeted advertising or enhances productivity. It can also be traded on various data marketplaces or shared between business partners. While user data contains useful information (e.g., purchasing habits) that unlock such added values, it can explicitly or implicitly contain users' private information (e.g., medical conditions). As data custodians, such businesses have both legal and ethical obligations to protect sensitive user information.  A research thrust has been designing provable mechanisms that transform such data to maintain its usefulness for a given task/analysis when shared with a third party, while minimizing the capability of such party to infer the sensitive information \cite{Dwork2011, PvsInfer2012, PF2014, Roy_ISIT}. Achieving a good balance between utility and privacy in shared data remains a challenging research topic and a barrier for businesses in trading/sharing more datasets.

One statistical privacy-preserving paradigm is based on information theory. It assumes the distribution between a sensitive variable $S$ and a non-sensitive variable $X$, denoted by $p(s,x)$,  is known or can be estimated. For instance, publicly available data links consumption of more than one soft drink per day (variable $X$) to increased likelihood of diabetes (variable $S$). The aim is to probabilistically or deterministically perturb $X$ into another variable $Y$ to be shared such that an information-theoretic measure of privacy leakage is minimized, while an acceptable utility from $Y$ can still be attained. Privacy leakage and utility can both be measured by the Shannon mutual information, denoted by $I(S;Y)$ and $I(X;Y)$ respectively, as in the privacy funnel \cite{PF2014, Ding2019ITW}. Recently, $\alpha$-Sibson generalization of mutual information \cite{verdu2015alpha}, denoted by $I_\alpha(S;Y)$, was proposed as another measure of privacy leakage \cite{sankar_alpha}, with maximal privacy leakage defined as $I_\infty(S;Y) $ \cite{issa2019operational}.
 
A stronger (non-average) measure of privacy is based on lift and log-lift \cite{PvsInfer2012, Watchdog2019}, which are defined as $l(s,y) =\frac{p(s,y)}{p(s)p(y)}$ and $i(s,y) = \log l(s,y)$, $s\in \S, y\in \Y$, respectively. Whenever the absolute value of log-lift (abs-log-lift for short) is upper bounded as $|i(s,y)| \leq \eps$ for all $s\in \S, y\in \Y$, we say that $\epsilon$-log-lift is attained between $S$ and $Y$.\footnote{A more accurate name is $\eps$-abs-log-lift, but for brevity we use $\eps$-log-lift.} Note that $I(S;Y) = \mathbb E[i(S;Y)]$ and  $I_\infty(S;Y) = \log\mathbb E [\max_{s\in \S}l(s,Y)]$. Therefore, $\epsilon$-log-lift is a strong condition on privacy, which guarantees both $I(S;Y) \leq \eps$ and $I_\infty(S;Y) \leq \eps$. In fact, it was shown in \cite{PvsInfer2012} that it also guarantees $2\epsilon$-differential privacy (DP) and in \cite{Watchdog2019} that it guarantees $I_\alpha(S;Y) \leq \frac{\alpha}{\alpha-1}\epsilon$, for $\alpha > 1$.

\subsection{Contributions}
A main research challenge is the design of an optimal privacy preserving mechanism $p(y|x)$ to achieve $\epsilon$-log-lift while maintaining the best utility for the released variable $Y$, which we measure by $I(X;Y)$. In \cite{Watchdog2019}, the authors started with a \emph{desired} bound $\epsilon$ on the abs-log-lift. Subset $\X_\eps \subseteq \X$ contains elements that already satisfy $|i(s,x)|\leq \eps$ for all $s\in \S$ and are shared unchanged. Elements in $\Xepsc = \X\setminus \X_\eps$ are perturbed into $y\in \Y$ through a probabilistic mapping $p(y|x)$. This scheme was referred to in \cite{Watchdog2019} as \emph{privacy watchdog.} An upper bound for the resultant abs-log-lift was given, but can often be loose. In this context, our paper addresses the following questions:
\begin{enumerate}
\item Given $\eps$, $\X_\eps$ and $\Xepsc$, what is the tightest bound on the resultant abs-log-lift for all pairs $(s,y)$ and what are the randomized mechanisms that achieve this tight bound?
\item What are the critical values of $\eps$  that affect $\X_\eps$, $\Xepsc$?
\item How should one choose a good value for $\eps$ with an overall low abs-log-lift and good utility $I(X;Y)$?
\end{enumerate}

In Section \ref{sec:suff:necc}, we address question 1. For a given $\eps'$, we provide a sufficient and necessary condition for $\eps'$-log-lift  to be attainable after a privacy-preserving randomization $p(y|x)$. From this result, we provide an explicit expression for the best attainable $\eps'$-log-lift and show that any $X$-invariant randomization achieves this optimum (e.g., merging all symbols in $\Xepsc$ into a single arbitrary element $y^* \in \Y$). In Section \ref{sec:critical}, we address question 2 and explore critical values of $\eps$ based on the joint distribution $p(s,x)$. In Section \ref{sec:PUT}, we address question 3 and show how a strict bound $\eps$ on abs-log-lift can severely damage utility. To address this, we propose a relaxed \emph{probabilistic} ($\eps$, $\delta$)-log-lift watchdog, where $\eps$-log-lift is guaranteed with probability $1-\delta$. We present a concrete heuristic ($\eps$, $\delta$)-log-lift algorithm which starts from the original watchdog bi-partition $\X_\eps$ and $\Xepsc$ and judiciously exempts some elements satisfying $(\eps,\delta)$-log-lift in $\Xepsc$ from randomization. The remaining elements in $\Xepsc$ go through a similar $X$-invariant randomization (such as merging all remaining elements). The algorithm admits an additional parameter $\bar\eps > \eps$ that guarantees abs-log-lift never exceeds $\bar\eps$. Numerical simulations show that such relaxation can significantly improve utility.

\section{System Model and Problem Motivation}\label{sec:model}
We consider the Markov chain $S \to X \to Y$, where $S \in \S$ represents some sensitive data, $X \in \X$ represents some non-sensitive data, and $Y \in \Y$ is shared with a third party for a given task or analysis requiring  some utility about $X$. The objective is to design a privacy-preserving randomized mechanism with conditional distribution $p(y|x)$, $x \in \X, y \in \Y$, such that an acceptable privacy protection for $S$ and utility level about $X$ can be simultaneously obtained.

Let us first understand the privacy leakage about $S$ if $X$ were to be released unperturbed. Denoting the joint probability $p(s,x)$, in previous work \cite{PvsInfer2012, Watchdog2019}, \emph{log-lift} was defined as:
\[i(s,x) \triangleq \log\left(\frac{p(s,x)}{p(s)p(x)}\right) = \log\left(\frac{p(x|s)}{p(x)}\right) = \log\left(\frac{p(s|x)}{p(s)}\right),\]
for $s \in \S, x \in \X$. In this definition, note the log-ratio measures the statistical distance of the posterior belief $p(s|x)$ on the sensitive data $s$, upon observation $x$, to the prior belief $p(s)$. A larger deviation of $p(s|x)$ from $p(s)$ will make the posterior belief more deterministic and thus more vulnerable to malicious estimation, guessing or any other type of inference. So, if this deviation is bounded, many statistical measures of  information privacy leakage will become bounded as a result, as discussed in the Introduction.

Together, the maximum and minimum value of log-lift over all $s,x$, or $\max_{s,x} |i(s,x)|$,  denote the privacy level: A smaller $\max_{s,x} |i(s,x)|$ indicates more privacy and the perfect privacy is attained if $\max_{s,x} |i(s,x)| = 0$, \emph{i.e.}, there is $0$-leakage or independence between $S$ and $X$.\footnote{This is also the sufficient condition for perfect DP in \cite{Dwork2011,LDP2014}.} To attain a desired $\epsilon$-log-lift by a privacy-preserving mechanism $p(y|x)$, a convenient method that was suggested in \cite{Watchdog2019}, is to bi-partition $\X$ into
\[\X_\eps= \{x\in \X: |i(s,x)| \leq \eps, \quad \forall s\in \S\},\]
and $\X_\eps^c = \X \setminus \X_\eps$. Elements in $x \in \X_\eps$ are directly published and those in $\Xepsc$ are randomized using a valid, but otherwise unspecified distribution $R(y)$ over $y\in \Xepsc$ such that $\sum_{y \in \Xepsc} R_Y(y)= 1$. Such choice $R(y)$ signifies invariance of mapping  to $x \in \Xepsc$. Overall, this results in
\begin{equation}  \label{eq:randomization1}
	 \begin{aligned}
	 		p(y|x)
	 		&=
	 			\begin{cases}
	 				 1_{\{x=y\}} & x,y \in \X_\epsilon, \\
					 					 R(y) & 			x,y \in \X_\epsilon^c,	 \\
	 				 0 	&\text{otherwise.}
	 			\end{cases}
	 \end{aligned}
\end{equation}
Even though the desired $\eps$-log-lift is guaranteed in $\X_\eps$, in general there is no guarantee that $\eps$-log-lift is also attained in $\Y$ after randomization $R(y)$. It is important to determine the worst-case abs-log-lift after randomization and the probability of it occurring. The authors in \cite{Watchdog2019} derived an upper bound on the resultant abs-log-lift for all pairs  $(s,y)$, $s \in \S, y \in \Xepsc$. However, this bound can be quite loose and may cause unnecessary alarm for data custodians prompting them to take unnecessary drastic actions to ``curb" the resultant (loose upper bound) on abs-log-lift. This might unnecessarily  damage the utility. In addition, it is not clear whether there is any advantage in optimizing \emph{general} privacy preserving
mappings $p(y|x)$ (that are not necessarily invariant to $X$) to achieve an optimum log-lift and/or utility. We will show in the following section (see Remark~\ref{rem1}) that the optimality of the log-lift in $\Xepsc$ can indeed be attained by any valid $R(y)$, invariant to $x \in \Xepsc$.

\begin{figure}[t]
\centering
\includegraphics[width=0.8\columnwidth]{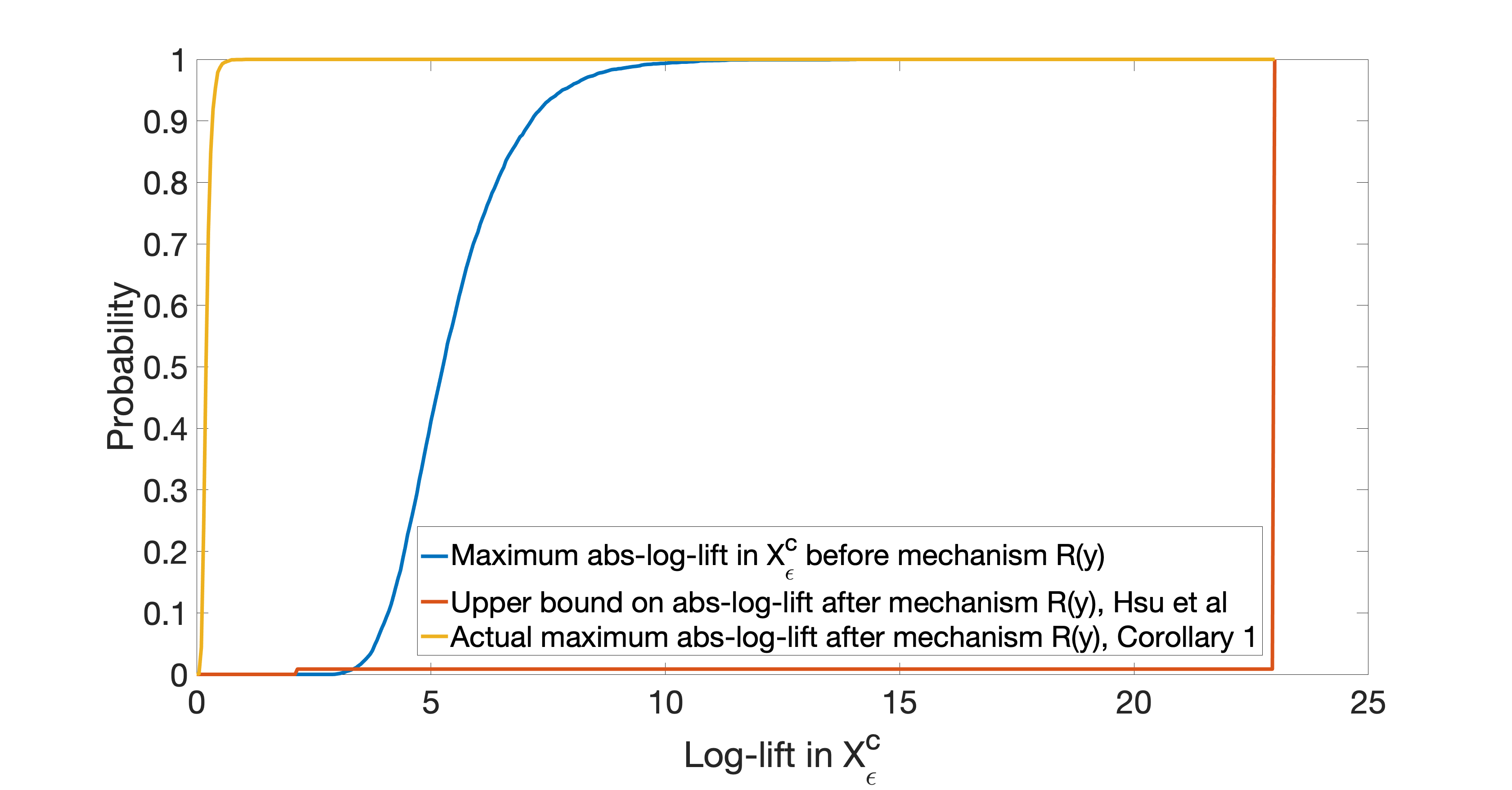}\caption{\small{Comparison of maximum abs-log-lift before and after randomization $R(y)$, (blue and yellow curves, respectively), as well as comparison with upper bound from \cite{Watchdog2019} (red curve). Here, $\eps = 2$, $|\X| = 20$ and $|\S| = 15$. For blue and red curves see Section \ref{sec:model} and for the yellow curve see Section \ref{sec:suff:necc}.}}\label{fig:bounds}
\end{figure}

Fig. \ref{fig:bounds} shows the former issue. The red curve is the cumulative  distribution function (CDF) of the upper bound on abs-log-lift in $\Xepsc$ for $\epsilon = 2$ according to \cite{Watchdog2019}[Proposition 2] after randomization $R(y)$ is applied to elements in $\Xepsc$ for 5000 randomly generated distributions $p(s,x)$, $s \in \S, x \in \X$. The upper bound actually tends to infinity almost all the time (in the figure it is capped to 23 for   illustration purposes). The original distribution of maximum abs-log-lift in $\Xepsc$ without any randomization is shown in the blue curve. It is clear that the upper bound  widely overestimates the maximum abs-log-lift. This is particularly the case for rather large $\eps$, which may prompt custodians to unnecessarily lower $\eps$ and expand $\Xepsc$, which as we will see in Section \ref{sec:PUT} will damage utility. This motivates the next section: to determine the best possible abs-log-lift in  $\Xepsc$ over all general randomizations $p(y|x)$.

\section{The best possible abs-log-lift in $\Xepsc$}\label{sec:suff:necc}
For the randomization scheme \eqref{eq:randomization1}, we have $p(y|s) = \sum_{x} p(y|x)p(x|s)$ due to the Markov property and $p(y) = \sum_x p(y|x) p(x)$ in the log-lift $\log \frac{p(y|s)}{p(y)}$.
Note, this randomization  preserves the original abs-log-lift in $\X_\eps$ to be at most $\epsilon$. Therefore, we focus on the abs-log-lift in $\X_\eps^c$.
We say that $p(y|x)$ in \eqref{eq:randomization1} attains $(\eps',\Xepsc)$-log-lift if
\[ |i(s,y)| = \Big| \log \frac{p(y|s)}{p(y) } \Big| \leq \eps', \quad \forall y \in \Xepsc, s \in \S.  \]
Generalizing the notion of log-lift from singular elements $x\in\X$ to subsets $\X_Q \subseteq \X$, we define
\begin{equation}\label{i:c}
i(s, \X_Q) \triangleq 	 \log \frac{p(\X_{Q}|s)}{p(\X_Q)},
\end{equation}
where $p(\X_Q) = \sum_{x \in \X_Q} p(x)$ and $p(\X_Q|s)  =  \sum_{x \in \X_Q} p(x|s)$. In particular, $i(s, \Xepsc)$ for $\X_Q = \Xepsc$ will be crucial in what follows, where we show both the achievability of $(\eps',\Xepsc)$-log-lift and the optimal value of $\eps'$ are determined by $i(s,\Xepsc)$.

\begin{theorem}\label{theorem:conditions}
There exists a valid randomization $p(y|x),  x,y \in \X_\epsilon^c$ that attains $(\epsilon',\X_\eps^c)$-log-lift
if and only if
\begin{equation} \label{eq:Condition1}
| i(s,\X_\epsilon^c) | \leq \eps', \quad  s \in \S.
\end{equation}
\end{theorem}
\begin{IEEEproof}
To prove the sufficient condition, consider the  privacy-preserving mechanism $p(y|x)$ such that for each $y \in \Xepsc$, all non-zero $p(y|x)$ have the same value. That is, $p(y|x) = R(y), \forall x \in \Xepsc$ as  in \eqref{eq:randomization1},  where $R(y)$ remains constant (invariant) in $x \in \Xepsc$, but obeys the probability constraint $\sum_{x \in \X_\eps^c} R(y) = 1$. We call this $p(y|x)$ an $X$-invariant randomization for $x,y\in \X_\eps^c$.
For each $s$, rewrite the condition \eqref{eq:Condition1} using \eqref{i:c} as
\[e^{-\epsilon'}p(\X_\epsilon^c)  \leq  p(\X_{\eps}^c|s) \leq  e^{\epsilon'}p(\X_\epsilon^c). \]
For each $s$, the following inequalties hold for all $y \in \X_\eps^c$. Multiply each side of the above by $R(y)$ to get
\begin{align*}
e^{-\epsilon'}R(y) p(\X_\epsilon^c) &\leq R(y)p(\X_\epsilon^c|s)  \leq  e^{\epsilon'}R(y) p(\X_\epsilon^c) \\
& \Rightarrow\\
e^{-\epsilon'} \hspace{-1.5mm}\sum_{x \in \X_\epsilon^c} p(y|x)p(x)  &\leq \hspace{-2mm}\sum_{x \in \X_\epsilon^c} p(y|x)p(x|s) \hspace{-1mm}\leq  e^{\epsilon'}\hspace{-1.5mm}\sum_{x \in \X_\epsilon^c}   p(y|x)p(x)\\
\Rightarrow\\
e^{-\epsilon'} p(y)  &\leq p(y|s)  \leq e^{\epsilon'} p(y).
\end{align*}

To prove the necessary condition, for each $s \in \S, y \in \Xepsc$, let us focus on the lower bound $-\eps'$ of $i(s,y)$, where we have
\begin{align*}
e^{-\epsilon'} p(y)  &\leq p(y|s) \\
& \Rightarrow\\
e^{-\epsilon'} \sum_{x \in \X_\epsilon^c} p(y|x)p(x)  &\leq \sum_{x \in \X_\epsilon^c} p(y|x)p(x|s).
\end{align*}
Summing both sides over all $y \in \Xepsc$ and rearranging, we get
\begin{align*}
e^{-\epsilon'} \sum_{x \in \X_\epsilon^c} p(x) \underbrace{\sum_{y \in \X_\epsilon^c}p(y|x)}_{=1} &\leq \sum_{x \in \X_\epsilon^c} p(x|s)\underbrace{\sum_{y \in \X_\epsilon^c}p(y|x)}_{=1},
\end{align*}
from which we conclude $e^{-\epsilon'}p(\X_\epsilon^c)  \leq  p(\X_{\eps}^c|s)$. The inequality $p(\X_{\eps}^c|s) \leq e^{\epsilon'}p(\X_\epsilon^c)$ for the upper bound $\eps$ on $i(s,y)$  can be similarly proved. 
In Appendix \ref{app:A}, we present a proof of the necessary condition by contradiction.
\end{IEEEproof}

For $\X_Q\subseteq \X$, define
\begin{equation}\label{eq:eps:gen}
\eps (\X_Q ) \triangleq \max_{s \in \S} |i(s, \X_Q)|,
\end{equation} where $i(s, \X_Q)$ was defined in  \eqref{i:c}.
Using Theorem~\ref{theorem:conditions}, we obtain the optimal/minimal abs-log-lift that is attainable in $\X_\eps^c$:

\begin{corollary} \label{coro1} For a given value of $\epsilon$, the minimum value of $\epsilon'$ such that $(\epsilon,\X_\eps^c)$-log-lift is attainable is given by
\begin{equation}\label{eps:c}
	\epsilon^c \doteq \eps (\Xepsc ) = \max_{s\in \S} |i(s, \X_{\eps}^c)|.
\end{equation}
Moreover, the best abs-log-lift $\eps^c$ is attained by any valid $X$-invariant probability distribution $R(y)$ that only depends on $y$, but remains constant for all $x \in \X_\eps^c$.
\end{corollary}
The yellow curve in Fig. \ref{fig:bounds} shows the best abs-log-lift $\epsilon^c$ in $\Xepsc$ \emph{after} randomization $R(y)$, which shows its efficacy in bringing the abs-log-lift to a much lower level than the original maximum log-lift \emph{before} randomization, thereby assuring data custodians that the overall abs-log-lift remains below $\eps$.

\begin{remark} \label{rem1}
As explained in the proof of Theorem~\ref{theorem:conditions}, in \eqref{eq:randomization1}, $R(y)$ refers to a privacy-preserving mechanism on $\Xepsc$ that only depends on $y$, but remains constant for all $x \in \X_\eps^c$. For example, the uniform distribution $R(y) = \frac{1}{|\X_\eps^c|}$ is one such candidate. It also contains as a special case, the merging solution, \emph{e.g.}, the method in \cite{Ding2019ITW,PF2014} when there is only one $y^* \in \Xepsc$ such that $R(y^*) = 1$ for all $x \in \X_\eps^c$ and $R(y) = 0$, otherwise. This $R(y)$ is exactly the one used in \cite{Watchdog2019}. However, the optimality of it was not shown in \cite{Watchdog2019}.

\end{remark}

In Appendix \ref{app:B}, we directly prove Corollary~\ref{coro1} without using Theorem~\ref{theorem:conditions}, which clearly shows that we cannot have $\eps' < \eps^c$ by any randomization $p(y|x)$ in $\X_\eps^c$ other than the $X$-invariant randomization for $x,y \in \X_\eps^c$ in \eqref{eq:randomization1}.

In the sense of attaining the best abs-log-lift in $\Xepsc$, $p(y|x)$ in \eqref{eq:randomization1} is the optimal solution. We will use \eqref{eq:randomization1} to study privacy-utility tradeoff in Section~\ref{sec:PUT}.

\section{Critical Values of $\eps$ and $\eps^c$}\label{sec:critical}

The interpretation of Corollary~\ref{coro1} is that the optimal bound $\epsilon'$ on abs-log-lift only depends on the original data statistics $p(s,x)$. That is, once the value of $\epsilon$ is given, $\eps^c$, the minimal attainable $\epsilon'$,  is known.
For finite alphabets $\S$ and $\X$, the set $\X_\eps$ is determined by $|\X|$ discrete/finite thresholds of $\epsilon$ as follows.

Recall from \eqref{eq:eps:gen}, $\eps(x) \triangleq \max_{s \in \S} |i(s,x)|$, which for singleton elements $x \in \X$,  is the maximum abs-log-lift across all $s\in \S$. Let $j = 1,\dotsc,|\X|$ be the ordering of $x\in \X$ such that $\epsilon(x_j)$ is (strictly) decreasing in $j$:
$$ \epsilon(x_1) > \epsilon(x_2) > \cdots > \epsilon(x_{|\X|}).$$
When the context is clear, we simplify the notation $\epsilon(x_j)$ to $\epsilon_j$ and refer to them as \emph{critical} $\eps_j$'s. We can verify that whenever $\epsilon \in [\epsilon_{j+1},\epsilon_j)$, $\X_\epsilon^c = \Set{x_1,\dotsc,x_j}$ and $\X_\eps = \X \setminus \Xepsc = \Set{x_{j+1},\dotsc,x_{|\X|}}$.
Then, $\X_\epsilon^c$ is characterized by the set chain:
\begin{equation}
	 \underbrace{\emptyset = \X_{\epsilon_0}^c}_{\epsilon > \epsilon_1} \subsetneq \X_{\epsilon_1}^c \subsetneq  \X_{\epsilon_2}^c \subsetneq \cdots \subsetneq \underbrace{\X_{\epsilon_{|\X|}}^c = \X}_{\epsilon < \epsilon_{|\X|} },
\end{equation}
where $\X_{\epsilon_{j+1}}^c \setminus \X_{\epsilon_{j}}^c = \Set{x_{j+1}}$. Note that for $\epsilon \in [\epsilon_{2},\epsilon_1)$, $\X_\epsilon^c = \Set{x_1}$ is singleton and therefore, it is not meaningful to design privacy preserving mechanisms for this case.

For a general bi-partition of $\X$ induced by $\X_Q$, where elements in $\X_Q$ are unchanged and elements in $\X_Q^c = \X \setminus \X_Q$ go through randomization, the effective abs-log-lift is
\[ \eps_{\text{eff}} (\X_Q) = \max \Set{ \max_{x \in \X_Q} \eps(x), \eps (\X_Q^c)}. \]
Denote $\eps_j^c =  \eps (\X_{\eps_j}^c)$ and note that $\eps_j^c$ can be larger than $\eps_j$. Based on Corollary~\ref{coro1}, for each $\eps_j$, the effective abs-log-lift over both $\X_{\eps_j}$ and $\X_{\eps_j}^c$ is $\eps_{\text{eff},j} = \max\{\eps_j,\eps_j^c\} $.

\section{Privacy-Utility Tradeoffs \& Relaxations}
\label{sec:PUT}

We study how the choice of $\epsilon$ as a bound on abs-log-lift affects the utility, measured by the mutual information $I(X;Y)$. Under the optimal randomization $p(y|x)$ in \eqref{eq:randomization1}, $p(y) = \sum_{x\in \Xepsc} p(y|x)p(x) = R(y)  \sum_{x\in \Xepsc} p(x) = R(y)p(\Xepsc)$ for all $y \in \X_\eps^c$ and the mutual information is given by
\begin{align}\label{eq:Hq1}
		I(X;Y) & = \sum_{x,y} p(x,y) \log \frac{p(x,y)}{p(x)p(y)}  \\\nonumber
				 & =  \hspace{-1mm} \sum_{x \in \X_\epsilon} \hspace{-1mm}p(x) \log \frac{1}{p(x)} \hspace{-1mm} +\hspace{-2mm}\sum_{y \in \X_\epsilon^c}\hspace{-0.5mm}\sum_{x \in \X_\epsilon^c} \hspace{-1mm} p(y|x)p(x) \log \frac{p(y|x)}{p(y)} \\\nonumber
				 & = \sum_{x \in \X_\epsilon} p(x) \log \frac{1}{p(x)}  \hspace{-1mm}+ \hspace{-2mm}\sum_{y \in \X_\epsilon^c} \hspace{-0.5mm}\sum_{x \in \X_\epsilon^c} \hspace{-1mm}  R(y) p(x) \log \frac{1}{p(\X_\epsilon^c)} \\\nonumber
				 & = H(X) + \sum_{x \in \X_\epsilon^c} p(x) \log \frac{p(x)}{p(\X_\epsilon^c)} \\
				 &= H(X) - p(\X_\epsilon^c) H(q)\label{eq:Hq11}
\end{align}
where in the last equality, $q(x) \triangleq \frac{p(x)}{p(\Xepsc)}$ is the re-normalized distribution of $X$ over $\Xepsc$ only. From \eqref{eq:Hq11}, it is clear that the larger the probability $p(\Xepsc)$ or the more uniform the distribution $q(x)$, the smaller the utility $I(X;Y)$ will be attained. For a given subset $\X_Q\subseteq \X$, define the normalized mutual information loss (NMIL) as
\begin{equation}\label{eq:NMIL}
\text{NMIL}(\X_Q) = \frac{p(\X_Q)H(q)}{H(X)},
\end{equation}
where $H(q) = -\sum_{x \in \X_Q} \frac{p(x)}{p(\X_Q)} \log \frac{p(x)}{p(\X_Q)}$ and NMIL$\in [0,1]$. It can be verified that NMIL is monotonic in $\X_Q$: NMIL$(\X_Q) < $ NMIL$(\X_{Q'})$ for $\X_Q \subsetneq \X_{Q'}$.
We calculate NMIL$(\Xepsc)$ over 5000 random generations of the joint probability distribution $p(s,x)$ with $|\X| = 20$ and $|\S| = 15$, from which we calculate both log-lift $i(s,x)$ and NMIL. Fig. \ref{fig:loss:eps:delta} shows the CDF of randomly obtained NMIL over these 5000 trials for two values $\eps = 1$ and $\eps = 2$ (here only two curves with $\delta = 0$ are relevant, where $\delta$ will be defined shortly). We observe from the rightmost curve that for a relatively strict privacy watchdog $\eps = 1$, a heavy price for the utility is paid where the NMIL is 0.7 or more almost all the time.
\vspace{-3mm}
\subsection{Introducing Probabilistic $(\eps, \delta)$-Log-lift Privacy}
As seen above, the condition of a desired bound $\eps$ on abs-log-lift being 100\% satisfied is quite strict and can lead to a substantial utility loss. Upon choosing $\epsilon \in [\epsilon_{j+1},\epsilon_j)$, both $\X_\eps = \{j+1, \cdots, |\X|\}$ and $\Xepsc = \{1, \cdots, j\}$ are fixed. However, this universal condition does not take into account the behavior of  the log-lift  across $s \in \S$ nor the joint probability $p(s,x)$. Note, $p(s,x)$ denotes the chances for a particular log-lift $i(s,x)$ to appear in the dataset. For example, it could very well be that for a given $x\in \X$, the abs-log-lift is very large for a particular pair $(s^*,x)$, but with a small probability $p(s^*,x)$ and very small for all other $(s,x)$ with a much larger probability $\sum_{s\neq s^*} p(s,x)$.\footnote{Note, if $\frac{p(s^*,x)}{\sum_{s} p(s,x)\sum_{x} p(s^*,x)}$ is very small, $|i(s^*,x)|$ will be very large.} In this part, we relax the condition of abs-log-lift from being 100\% (or `universally') satisfied to being satisfied with probability $1-\delta$ for a small slack parameter $\delta$. This relaxation is a counterpart of \emph{probabilistic} ($\epsilon$, $\delta$)-DP \cite{Meiser2018ApproximateAP, ProbDP}, where $\eps$-DP is guaranteed with probability $1-\delta$.\footnote{We emphasize the distinction between approximate ($\eps$, $\delta$)-DP, which is more prevalent in the DP literature, and probabilistic ($\eps$, $\delta$)-DP. For a treatment of subtitles among different relaxations of DP see \cite{Meiser2018ApproximateAP}. Our relaxation scheme is inspired by the probabilistic relaxation of DP, not approximate ($\eps$, $\delta$)-DP.}
Recall the definition of $i(s,\X_Q)$ in \eqref{i:c}. Now
define
\begin{equation}\label{eq:delta:eps}
\Delta(\eps, \X_Q)\triangleq \sum_{s\in \S: |i(s,\X_Q)| > \eps}p(s,\X_Q).
\end{equation}
For non-singleton $\X_Q$, $\Delta(\eps, \X_Q)$ measures the probability that the abs-log-lift goes above $\eps$ if we randomize $\X_Q$ according to $R(y)$ in \eqref{eq:randomization1}. For singleton  $\X_Q = \{x\}$ for some $x\in \X$, $\Delta(\eps, x)$ measures the probability that the abs-log-lift is above $\eps$ for an unperturbed $x$. Define the overall probability of breaching $\eps$-log-lift across the bi-partition of $\X$ induced by $\X_Q$ as
\begin{equation}
\Delta_{\text{total}}(\eps, \X_Q) = \sum_{x \in \X_Q} \Delta(\eps, x) +  \Delta(\eps,\X_Q^c),
\end{equation}
where the first term signifies the probability of breaching the $\eps$-log-lift if we kept the elements in $\X_Q$ unperturbed. 
\begin{definition}\label{def:eps:delta}
Fix $\eps > 0$ and $0< \delta < 1$. If a bi-partition $(\X_Q, \X_Q^c)$ satisfies
\begin{equation}\label{eq:delatdef}
 \Delta_{\text{total}}(\eps, \X_Q) \leq \delta,
\end{equation}
we say it is a $\delta$-approximation $\eps$-log-lift partition on $\X$, or an ($\eps$, $\delta$)-partition for short, that achieves ($\eps$, $\delta$)-log-lift privacy.
\end{definition}

Let $\Gamma(\eps, \delta)$ denote the set of all  ($\eps$, $\delta$)-partitions on $\X$. We propose the following privacy-utility optimization.

\begin{proposition}\label{eq:prop:delta:opt}
Fix $\eps > 0$ and $\delta > 0$. The best mutual information utility $I(X;Y)$ subject to ($\eps$, $\delta$)-log-lift  privacy is the solution to the following problem:
\begin{equation}\label{eq:NMIL}
\min_{\X_Q \in \Gamma(\eps, \delta) } \text{NMIL}(\X_Q^c).
\end{equation}

\end{proposition}
In the absence of any known structure for $\Gamma(\eps, \delta)$, the optimization in Proposition \ref{eq:prop:delta:opt} seems combinatorial in nature. However, it is possible to limit the search in a systematic way to a subset of  $\Gamma(\eps, \delta)$ to find good bi-partitions  $(\X_Q, \X_Q^c)$  that can strike a better balance between $(\eps, \delta)$-log-lift privacy on the one hand and utility on the other.

\subsection{A Heuristic Algorithm and Numerical Simulations}
\label{sec:Exp}
   
The idea of the heuristic algorithm is as follows.  For a given $\eps$, we first determine the original $\X_\eps$ and $\Xepsc$ and then bootstrap from this bi-partition to judiciously remove some elements from $\Xepsc$ and add them back into $\X_\eps$ without violating the overall ($\eps$, $\delta$)-log-lift  privacy condition in Definition \ref{def:eps:delta}.  Note that $\delta_0 = \Delta_{\text{total}}(\eps,\Xepsc)$ 
as defined in \eqref{eq:delta:eps} is the probability that the  abs-log-lift $|i(s,\Xepsc)|$ is greater than $\eps$ to begin with. Therefore, relaxation of log-lift occurs when $\delta > \delta_0$.

Other than the three basic inputs $p(s,x)$, $\eps$ and $\delta$, the algorithm admits one additional parameter $\bar\eps$. If we wish to have no abs-log-lift greater than a given parameter $\bar{\eps} > \eps$ all the time, we can exclude those elements from being moved back to $\X_\eps$.  By setting $\bar{\eps}= \infty$, no upper limit is imposed on the resulting log-lift. To avoid unnecessary iterations in the algorithm, we refine $\X_\eps^c$ at  initialization as follows
\begin{equation} \label{eq:RefinedXepsc}
				\mathcal M = \Xepsc \setminus \Big\{ x \in \Xepsc \colon \delta(\eps, x) > \delta
				\text{ or } \eps(x)> \bar{\eps} \Big\},
\end{equation}
which removes all $x \in \X_\eps^c$ that incur $\delta(\eps, x) > \delta$, since moving them to $\X_\eps$ breaches the $\delta$ condition, and all $x \in \X_\eps^c$ that have $|i(s,x)| > \bar{\eps}$ for some $s\in \S$, since moving them to $\X_\eps$ incurs maximum abs-log-lift greater than $\bar{\eps}$. The algorithm then sorts elements in $\mathcal M= \Set{x_1, x_2, \cdots}$ such that $\delta(\eps, x_i) \leq \delta(\eps, x_j)$ for $i \leq j$. It then greedily adds elements from $\mathcal M$ back to $\X_\eps$ in increasing order of $\delta(\eps, x)$, as long as $\delta$ and $\bar\eps$ conditions are not violated and the utility measured by NMIL is improved.

\begin{figure}[t]
\includegraphics[width=0.9\columnwidth]{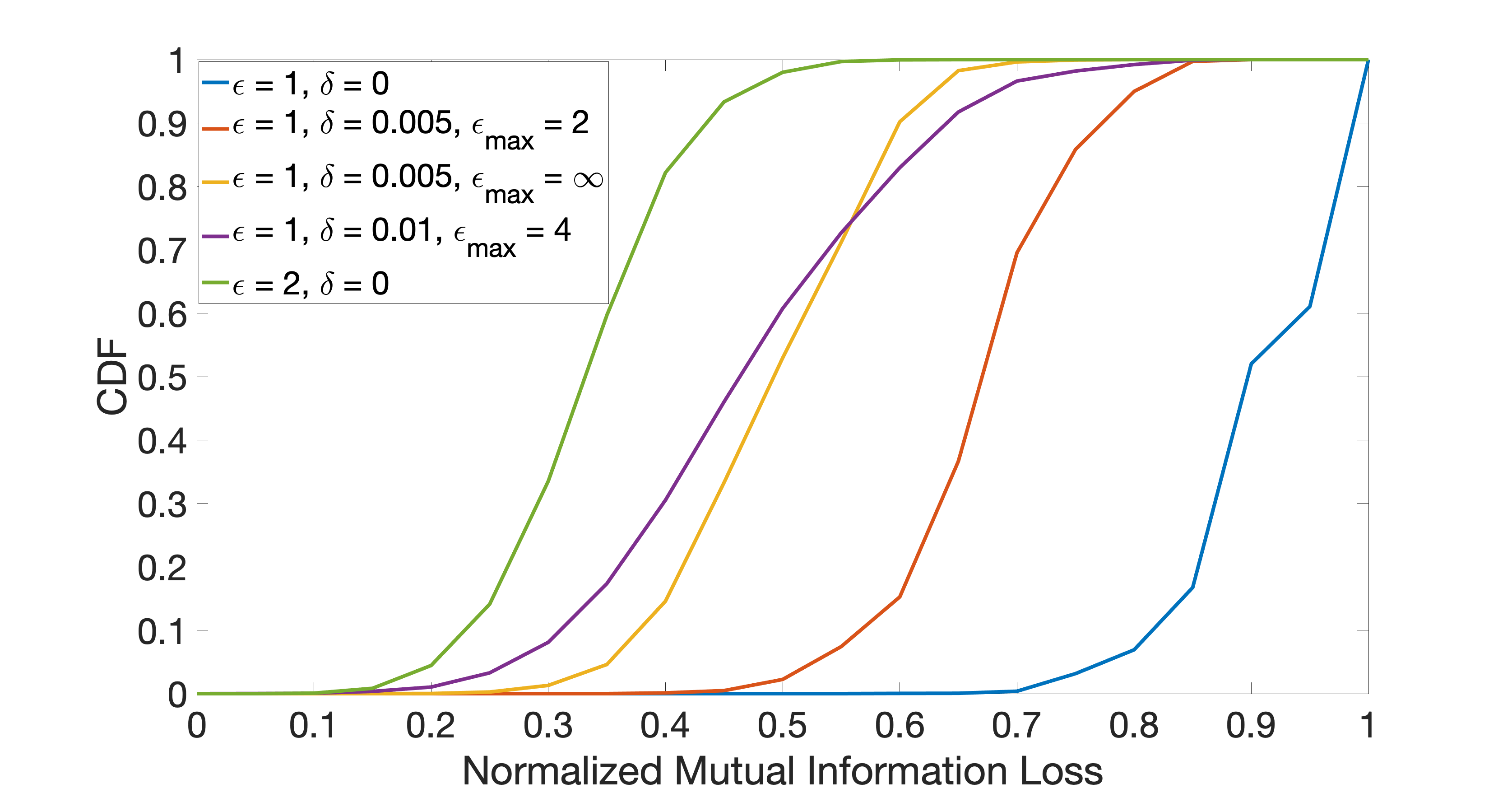}\caption{NMIL for $\eps = 1, 2$ and different $\delta$ and $\eps_{\max} = \bar\eps$.}\label{fig:loss:eps:delta}
\end{figure}

Fig.~\ref{fig:loss:eps:delta} compares the CDF of NMIL for numerical simulations of Algorithm 1 against the original non-relaxed cases. We set a global value for $\eps = 1$ and three different relaxations as follows. In the first setting, we set $\delta = 0.005$ and $\bar\eps = 2$. In the second setting, $\delta = 0.005$ and $\bar\eps = \infty$ ($\bar\eps = 1000$ in the simulations). In the third setting, we further relax $\delta = 0.01$ with $\bar\eps = 4$.\footnote{In our simulations, we always observed $\eps^c = \max_{s \in \S} |i(s, \Xepsc)| < \eps$, meaning $\delta_0 = 0$. Therefore, we could set any global $\delta > 0$ in our experiments. Please refer to Appendix \ref{app:delta0} for more details.} Compared to $\eps = 1$ and $\delta = 0$ and even with a very small relaxation $\delta = 0.005$ and $\bar\eps = 2$, we can improve the NMIL noticeably. But perhaps most interestingly, substantial gain in utility is observed when we further relax either $\bar\eps$ or $\delta$. In the yellow curve, we allow any abs-log-lift to happen: $\bar\eps = \infty$, but with maximum probability $\delta = 0.005$. Through this relaxation, we can get very close to the utility curve with $\eps = 2$ and strict $\delta = 0$ (purple curve) and even outperform it at high utility losses, while still ensuring $\eps = 1$ for 99.5\% of the time. In the green curve, we allow $\bar\eps = 4$ with probability $\delta = 0.01$, through which, we can substantially improve the utility curve compared to $\eps = 2$ and strict $\delta = 0$, while still ensuring $\eps = 1$ for 99\% of the time. Under this last setting, utility loss is at most 50\% more than 97\% of the time.

\section{Conclusion}
In this paper, we derived a tight performance bound on the abs-log-lift in the privacy watchdog framework. We showed that this bound is achievable via simple symbol merging or uniform randomiztion. We introduced a relaxation parameter $\delta$ into the $\eps$-log-lift privacy framework, which allows abs-log-lift to go over $\eps$ with probability at most $1-\delta$. We proposed a greedy algorithm to implement this idea and showed it can provide a better privacy-utility tradeoff compared to the original strict $\eps$-log-lift watchdog scheme. For future work, it will be interesting to incorporate the distribution $p(s,x)$ into other privacy measures to relax them into probabilistic  ($\eps$,$\delta$) measures. It will be also interesting to study the privacy watchdog scheme under \emph{full data} randomization \cite{ishwar:2017} where both $s\in \S$ and $x\in \X$ are used in the randomization $p(y|x,s)$. One can also pursue practical applications of the ($\eps$,$\delta$)-log-lift privacy scheme where data is either longitudal (has repeated or correlated entries) or is continuously updated and released.

    \begin{algorithm} [t] 
	       \small
	       \SetAlgoLined
	       \SetKwInOut{Input}{input}\SetKwInOut{Output}{output}
	       \SetKwFor{For}{for}{do}{endfor}
            \SetKwRepeat{Repeat}{repeat}{until}
            \SetKwIF{If}{ElseIf}{Else}{if}{then}{else if}{else}{endif}
	       \BlankLine
           \Input{$p(s,x)$ for all $s \in \S, x \in \X$; $\eps, \delta >\delta_0$; $\bar{\eps}$ where $\bar{\eps} > \eps$. }
	       \Output{($\eps$, $\delta$)-partition $(\X_{Q}, \X_{Q}^c)$.}
	       \BlankLine
                 Initialize: Obtain $\mathcal{M}$ from \eqref{eq:RefinedXepsc}; Sort $\mathcal{M} = \Set{x_1,x_2, \cdots}$ in increasing order of $\delta(\eps,x)$; NMIL $\leftarrow$ NMIL$(\Xepsc )$, $\X_Q \leftarrow \X_\eps$ and $\X_Q^c = \Xepsc$\;
                \For {$j= 1$ to $|\mathcal{M}|$} {
   \lIf{NMIL$(\X_{Q}^c \setminus \Set{x_j})<$ NMIL, $\delta_{\text{total}}(\eps, \X_{Q} \cup \Set{x_j}) \leq \delta$ and $\eps_{\text{eff}}(\X_Q\cup\Set{x_j}) \leq \bar \eps$}{NMIL $\leftarrow$ NMIL$(\X_{Q}^{c} \setminus \Set{x_j})$, $\X_{Q} \leftarrow \X_{Q} \cup \Set{x_j}$ and $\X_{Q}^c \leftarrow \X_{Q}^c \setminus \Set{x_j}$}
}
	   \caption{Greedy ($\eps$, $\delta$)-partitioning.}
	   \end{algorithm} \label{algo:heu}
	   
\bibliographystyle{IEEEtran}
\bibliography{PrivacyBIB}

\appendices

\section{Supplement Proof of Theorem~\ref{theorem:conditions}}\label{app:A}
We prove the necessary condition of Theorem~\ref{theorem:conditions} by contradiction. 
Assume that there exists a randomization $p(y|x)$ that attains $(\epsilon',\Xepsc)$-log-lift for all $y \in \X_\epsilon^c$ when \eqref{eq:Condition1} does not hold. That is, either $p(\Xepsc|s) >  e^{\epsilon'} p(\Xepsc)$ or $p(\Xepsc|s) <  e^{-\epsilon'} p(\Xepsc) $. Assume the first case $p(\Xepsc|s) >  e^{\epsilon'} p(\Xepsc)$ holds for some $s$. 

Pick any $\hat{y} \in \X_\epsilon^c$. Due to the $(\epsilon', \Xepsc)$-log-lift assumption 
\begin{align*}
			& \sum_{x \in \X_\epsilon^c} \big( 1-\sum_{y \in \X_\epsilon^c \colon y \neq \hat{y}} p(y|x) \big) p(x|s) \\
			& = \sum_{x \in \X_\epsilon^c}  p(\hat{y}|x) p(x|s)  = p(\hat{y}|s) \leq p(\hat{y}) e^{\epsilon'}  = \sum_{x \in \X_\epsilon^c} p(\hat{y}|x) p(x) e^{\epsilon'} \\
			& = \sum_{x \in \X_\epsilon^c} \big( 1-\sum_{y \in \X_\epsilon^c \colon y \neq \hat{y}} p(y|x) \big) p(x) e^{\epsilon'}, 
\end{align*}
which is equivalent to 
\begin{multline*}
	 p(\Xepsc|s) -p(\Xepsc) e^{\epsilon'}  \leq \\
	 \sum_{y \in \X_\epsilon^c \colon y \neq \hat{y}}  \sum_x p(y|x)p(x) - p(x) e^{\epsilon'} \leq 0.
\end{multline*}
This contradicts the condition $ p(\Xepsc|s) >  e^{\epsilon'} p(\Xepsc)$. Note that this contradiction happens for each $(\epsilon',\Xepsc)$-log-lift randomization $p(y|x)$ and also can be shown in the same way when the second case $p(\Xepsc|s) <  e^{-\epsilon'} p(\Xepsc) $ holds. Therefore, there must not exist a randomization $p(y|x)$ that attains $(\epsilon',\Xepsc)$-log-lift for any $y \in \X_\epsilon^c$ when \eqref{eq:Condition1} does not hold. \hfill \IEEEQED

\section{Complete Proof of Corollary~\ref{coro1}}\label{app:B}

Let $\bar{s} \in \argmax p(\X_\epsilon^c|s)$ and $\underline{s} \in \argmin p(\X_\epsilon^c|s)$. We prove Corollary~\ref{coro1} by contradiction. Assume there exists another $p(y|x)$ that attains a lift strictly smaller than $\eps^c$. That is, both
\begin{align}
		& \frac{\sum_{x \in \X_\epsilon^c}p(y|x)p(x|s)}{\sum_{x \in \X_\epsilon^c}p(y|x)p(x)}  <  i(\bar{s},\Xepsc) = \frac{p(\X_\epsilon^c | \bar{s} )}{p(\X_\epsilon^c)},   \label{eq:UniOptAux1}\\\nonumber
		&\text{and}\\
		& \frac{\sum_{x \in \X_\epsilon^c}p(y|x)p(x|s)}{\sum_{x \in \X_\epsilon^c}p(y|x)p(x)}  > i(\underline{s},\Xepsc) = \frac{p(\X_\epsilon^c|\underline{s})}{p(\X_\epsilon^c)}  \label{eq:UniOptAux2}
\end{align}
hold for all $y \in \Xepsc$ and $s \in \S$. 
But, for any $y$, we have  
\small{
\begin{align} 
		 & \quad \frac{\sum_{x \in \X_\epsilon^c}p(y|x)p(x|s)}{\sum_{x \in \X_\epsilon^c}p(y|x)p(x)}  - \frac{p(\X_\epsilon^c|\bar{s})}{p(\X_\epsilon^c)}  \nonumber \\
		 & = \frac{p(\X_\epsilon^c) \sum_{x \in \X_\epsilon^c}p(y|x)p(x|s) - p(\X_\epsilon^c | \bar{s}) \sum_{x \in \X_\epsilon^c}p(y|x)p(x)}{p(\X_\epsilon^c) \sum_{x \in \X_\epsilon^c}p(y|x)p(x)} \nonumber \\
		 & = \frac{ \sum_{x \in \X_\epsilon^c} p(y|x) \big(  p(\X_\epsilon^c) p(x|s) - p(\X_\epsilon^c | \bar{s}) p(x) \big) }{p(\X_\epsilon^c) \sum_{x \in \X_\epsilon^c}p(y|x)p(x)}  \label{eq:IneqAux2}  \\
		  & = \frac{ \sum_{x \in \X_\epsilon^c} (1- \sum_{y' \in \X_\epsilon^c \colon y' \neq y}p(y'|x)) \big(  p(\X_\epsilon^c) p(x|s) - p(\X_\epsilon^c | \bar{s}) p(x) \big) }{p(\X_\epsilon^c) \sum_{x \in \X_\epsilon^c}p(y|x)p(x)}  \nonumber \\
		  & > \frac{ \sum_{x \in \X_\epsilon^c} \big( p(\X_\epsilon^c) p(x|s) - p(\X_\epsilon^c | \bar{s}) p(x) \big) }{p(\X_\epsilon^c) \sum_{x \in \X_\epsilon^c}p(y|x)p(x)} \label{eq:IneqAux1} \\
		  & = \frac{ p(\X_\epsilon^c | s) - p(\X_\epsilon^c | \bar{s}) }{\sum_{x \in \X_\epsilon^c}p(y|x)p(x)}.  \label{eq:IneqAux}
\end{align}}
Here, the inequality \eqref{eq:IneqAux1} is because, for all $y' \in \X_\epsilon^c \colon y' \neq y$,  \eqref{eq:UniOptAux1} holds and therefore 
$ \sum_{x \in \X_\epsilon^c} p(y'|x) \big(  p(\X_\epsilon^c) p(x|s) - p(\X_\epsilon^c | \bar{s}) p(x) \big) < 0$. 
For $s = \bar{s}$, we have \eqref{eq:IneqAux} = 0, i.e., \eqref{eq:UniOptAux1}  does not hold, which is a contradiction.
To show the contradiction in \eqref{eq:UniOptAux2}, consider the numerator of \eqref{eq:IneqAux2} and replace $\bar{s}$ with $\underline{s}$. We have
\begin{equation}
	\begin{aligned}
		 & \sum_{x \in \X_\epsilon^c} p(y|x) \big(  p(\X_\epsilon^c) p(x|s) - p(\X_\epsilon^c | \underline{s}) p(x) \big) \\
		 & \sum_{x \in \X_\epsilon^c} (1- \sum_{y' \in \X_\epsilon^c \colon y' \neq y}p(y'|x)) \big(  p(\X_\epsilon^c) p(x|s) - p(\X_\epsilon^c | \underline{s}) p(x) \big) \\
		 & <  \sum_{x \in \X_\epsilon^c} \big( p(\X_\epsilon^c) p(x|s) - p(\X_\epsilon^c | \underline{s}) p(x) \big)\\
		 & = p(\X_\epsilon^c) \big( p(\X_\epsilon^c | s) - p(\X_\epsilon^c | \underline{s}) \big)
	\end{aligned}
\end{equation}
which contradicts \eqref{eq:UniOptAux2} for $s = \underline{s}$. Therefore, Corollary~\ref{coro1}  holds. \hfill \IEEEQED

\section{Remarks on the relation of $\eps^c$, $\eps$ and $\delta_0$}\label{app:delta0}
For the log-lift $i(s,\Xepsc) = \log \frac{p(\Xepsc|s)}{p(\Xepsc)} $, apply the log-sum inequality to get
\begin{equation} \label{eq:BoundAux}
\sum_{x \in \Xepsc} \frac{p(x)}{p(\Xepsc)} i(s,x)   \leq i(s,\Xepsc) \leq \sum_{x \in \Xepsc} \frac{p(x|s)}{p(\Xepsc|s)} i(s,x). 
\end{equation}
Note that in \eqref{eq:BoundAux} the upper and lower bounds on $i(s,\Xepsc)$ are both in the form of convex combination of the log-lift $i(s,x)$ for instances of $s$ and $x \in \Xepsc$. 
We know that for each $x \in \Xepsc$, $\hat{\mathcal{S}}(x) = \Set{ s  \colon |i(s,x)| > \eps} \neq \emptyset$. Here,  $\hat{\mathcal{S}}(x)$ varies with $x$.  
In the LHS and RHS of \eqref{eq:BoundAux} consider any $x\in \Xepsc$. If for all $x' \in \Xepsc$ such that $x' \neq x$, we have $\hat{\mathcal{S}}(x) \cap \hat{\mathcal{S}}(x') = \emptyset $, it is possible to have $\sum_{x \in \Xepsc} \frac{p(x)}{p(\Xepsc)} i(s,x) > -\eps $ and $\sum_{x \in \Xepsc} \frac{p(x|s)}{p(\Xepsc|s)} i(s,x) < \eps$ so that\footnote{There might be other cases in which  \eqref{eq:StrBetter} holds. For example, if there are only two $x,x' \in \Xepsc$ such that $s \in \hat{\mathcal{S}} (x) \cap \hat{\mathcal{S}} (x') $, but $i(s,x) >\eps$ and $i(s,x') < -\eps$. } 
\begin{equation} \label{eq:StrBetter}
	\eps^c = \max_{s \in \S} |i(s, \Xepsc)| < \eps. 
\end{equation}
Intuitively, this happens when the elements of $s$ are such that a large abs-log-lift for one $s$ is compensated by a small abs-log-lift of another $s'$. If \eqref{eq:StrBetter} holds, then 
\begin{equation}
\begin{aligned}
\delta_0 & = \Delta_{\text{total}}(\eps,\Xepsc) \\
& = \sum_{s\in \S, x \in \X_\eps \colon  |i(s,x)| > \eps}p(s,x) + \sum_{s\in \S \colon  |i(s,\Xepsc)| > \eps}p(s,\Xepsc) \\
& = 0.
\end{aligned}
\end{equation}
When $\delta_0  = 0 $, all $\delta > \delta_0 = 0$ are valid relaxation choices, as was the case in the experiments in Section~\ref{sec:Exp}.
 
\end{document}